\documentclass{article}%
\usepackage{amsmath}
\usepackage{amsfonts}
\usepackage{amssymb}
\usepackage{graphicx}%
\providecommand{\U}[1]{\protect\rule{.1in}{.1in}}

\begin{document}

\begin{center}
\textbf{A\ NONSINGULAR\ BRANS WORMHOLE:\ AN\ ANALOGUE TO\ NAKED\ BLACK\ HOLES
}

\bigskip

\bigskip

\textbf{Amrita Bhattacharya,}$^{1,a}$\textbf{ Ramil Izmailov,}$^{2,b}$
\textbf{Ettore Laserra}$^{3,c}$ \textbf{and Kamal K. Nandi}$^{1,2,4,d}$

$\bigskip$

$^{1}$Department of Mathematics, University of North Bengal, Raja Rammohunpur,
Siliguri 734 013, India

$^{2}$Department of Theoretical Physics, Sterlitamak State Pedagogical
Academy, 50, Lenin Prospect, Sterlitamak 453103, Russia

$^{3}$DMI,\ Universit\`{a} di Salerno, Via Ponte Don Melillo, Fisciano 84084,
Salerno, Italy

$^{4}$Joint Research Laboratory, Bashkir State Pedagogical University, 3A,
October Revolution Street, Ufa 450000, Russia
\end{center}

\bigskip

\bigskip


\begin{center}

$^{a}$E-mail: amrita\_852003@yahoo.co.in

$^{b}$E-mail: izmailov.ramil@gmail.com

$^{c}$E-mail: elaserra@unisa.it

$^{d}$E-mail: kamalnandi1952@yahoo.co.in
\end{center}

\bigskip

\begin{center}
---------------------------------------------------------------------------

\textbf{Abstract}
\end{center}

In a recent paper, we showed the Jordan frame vacuum Brans Class I solution
provided a wormhole analogue to Horowitz-Ross naked black hole in the wormhole
range $-3/2<\omega<-4/3$. Thereafter, the solution has been criticized by some
authors that, because of the presence of singularity in that solution within
this range, a wormhole interpretation of it is untenable. While the criticism
is correct, we show here that (i) a singularity-free wormhole can actually be
obtained from Class I solution by performing a kind of Wick rotation on it,
resulting into what Brans listed as his independent Class II solution (ii) the
Class II solution has all the necessary properties of a regular wormhole in a
revised range $-2<\omega<-3/2$ and finally, (iii) naked black holes, as
described by Horowitz and Ross, are spacetimes where the tidal forces attain
their maxima above the black hole horizon. We show that in the non-singular
Class II spacetime this maxima is attained above the throat and thus can be
treated as a wormhole analogue. Some related issues are also addressed.

\begin{center}
---------------------------------------------------------------------------
\end{center}

\textbf{I. Introduction}

Lorentzian wormholes as possible astrophysical objects has been under active
investigation for quite some time now. In particular, the possibility of
occurrence of such objects in the Brans-Dicke theory is quite welcome since it
is a natural theory that emerged as a Machian alternative\footnote{Since the
Brans-Dicke theory contains a huge class of vacuum solutions, it might appear
that the properties of the space-time would not be dictated by the presence of
matter so that the theory is no longer Machian. This is not the case. The
Brans-Dicke scalar $\varphi$ is always sourced by the matter stress scalar
$T$, via $(\varphi^{;\rho})_{;\rho}=\frac{T}{2\omega+3}$. What we are doing is
that we are focusing only on the \textit{matter-free region }($T=0$), where
still $G\sim\left\langle \varphi\right\rangle ^{-1}$ at all space-time points.
The term in vogue, namely, \textquotedblleft vacuum Brans-Dicke theory" is a
misnomer since the exterior non-trivial fields ($g_{\mu\nu}$, $\varphi$) are
pretty much dictated by ponderable matter. It should more properly be called
the \textquotedblleft matter-free Brans-Dicke theory". Henceforth, by vacuum
we would mean matter-free. We thank an anonymous referee for raising this
issue.} to Einstein's theory of general relativity. To our knowledge, regular
wormhole properties of Brans-Dicke solutions (actually, what is referred to as
Class II below) are known for all $\omega<-3/2$ since 1973 from the work of
Bronnikov [22]. Then, Agnese and La Camera [1] have shown that the Brans-Dicke
scalar $\varphi$ can play the role of exotic matter provided the coupling
parameter $\omega<-2$. These are followed by the work of Visser and Hochberg
[2], and by the works in other classes of Brans solutions in the Jordan
Frame(JF) as well as their variants in the conformally rescaled Einstein Frame
(EF) [3,4]. The chronology stated here since the first work of Bronnikov are
certainly not exhaustive but there exist many interesting articles on
Brans-Dicke wormholes today (see, for instance [5-15]).

For static spherically symmetric Brans Class I solution in the $\omega=$
constant vacuum Brans-Dicke theory, we had earlier proposed a wormhole range
$-3/2<\omega<-4/3$ in the JF [4] and recently [15] developed a wormhole
analogy to Horowitz-Ross naked black holes [16] for $\omega<-2$. The latter is
the wormhole range obtained previously by Agnese and La Camera [1]. It is long
known that the Class I solution (which is inversion invariant under
$r\rightarrow\frac{B^{2}}{r}$) is plagued by curvature singularity: The $r=0$
flat spatial infinity is divided from the one at $r=\infty$ by a curvature
singularity at $r=B$. (The same singularity appears in its EF version too,
which is the Buchdahl solution [17]). So, the inversion transformation relates
two separately singular space-times and \textit{not} two regions of one
connected spacetime\footnote{We thank Professor Starobinsky for pointing this
out. The Class I solution has been criticized on the basis of a "no-ghost,
no-wormhole" theorem in Ref.[18] (see also [19]). However, as an aside, it
might be noted that Quiros, Bonal and Cardenas [20] have shown that the
cosmological singularity occurring in the EF theory is removed in the JF
precisely in the range $-3/2<\omega\leq-4/3$. Quiros [21] \ discusses the
impact of their result on the status of quantum cosmology in EF. Of course,
cosmology is not our concern here.}. Because of this, the Class I solution
cannot be interpreted as wormhole at all, which is certainly correct, and we
no longer stress on the Class I solution as a wormhole.

On the other hand, in view of the importance of Brans-Dicke theory in the
competing interpretation of various astrophysical phenomena, it is important
that a non-singular static spherically symmetric traversable wormhole solution
in the vacuum theory be found, and the analogy in the title be confirmed. The
purpose of the present paper is to achieve these.

In this direction, it was shown earlier [6] that the singular Buchdahl
solution in the EF could be Wick rotated into a new solution\footnote{We are
using the term \textquotedblleft Wick rotation\textquotedblright\ [as in
Eq.(24) below] rather loosely, which is used in quantum field theory in a very
specific situation.}, which was\ then identified as just the non-singular
Ellis Class III wormhole [22] in the EF. In the same spirit, one would expect
that the Brans Class I solution in the JF could likewise be Wick rotated into
a new non-singular wormhole in the JF. This is indeed possible. The results of
this paper show that the new solution (i) is just what Brans [23] listed as
his independent Class II solution\footnote{Since the transition Brans I
$\rightarrow$ II will be shown to be algebraically possible, \textit{albeit}
not by ordinary coordinate transformations, the two classes of solutions
cannot really be treated as independent, contrary to the common belief. This
fact was noted first by Bhadra and Sarkar [24] on the basis of the
complementary transition Brans II $\rightarrow$ I.} (ii) is non-singular
having all the desirable properties of a traversable wormhole but only in the
revised range $-2<\omega<-3/2$ and (iii) confirms the wormhole analogy to
naked black holes.

We point out that the main thread of development in this paper keeps only to
Brans wormholes in JF, while other useful questions have been mainly addressed
in the footnotes and in Sec.II, where we discuss actions in different
conformal frames. In Sec. III, we outline for wider readership the salient
features of a wormhole, while in Sec.IV we briefly review the singular Brans
Class I solution. This solution is then used to derive a non-singular solution
in Sec.V and its wormhole properties are enumerated in Sec. VI. This is
followed by the argument of wormhole analogy to naked black holes in Sec. VII.
Finally, Sec. VIII summarizes the contents. In the appendix, we derive
wormhole solutions in the EF. We take $8\pi G=c=1$ and a signature convention
$(-,+,+,+)$.

\textbf{II. The actions}

It would be useful to see how the vacuum Brans-Dicke action in JF relates to
that in the string frame and in the conformally rescaled EF\footnote{Two
anonymous referees, respectively, have commented on the string and conformal
related aspects of the Brans-Dicke theory. We address those comments here.}.
Hence, we start from the 4-dimensional, low energy effective action of
heterotic string theory compactified on a 6-torus. The tree level string
effective action in the string frame (SF or just S-frame), keeping only linear
terms in the string tension $\alpha^{\prime}$ and in the curvature
$\widetilde{\mathbf{R}}$, takes the following form in the matter free region
($S_{\text{matter}}=0$):
\begin{equation}
S_{\text{SF}}=\frac{1}{{\alpha^{\prime}}}\int{d^{4}x\sqrt{-\tilde{g}%
}e^{-2\tilde{\Phi}}}\left[  \widetilde{\mathbf{R}}{+4\tilde{g}^{\mu\nu}%
\tilde{\Phi}_{,\mu}\tilde{\Phi}_{,\nu}}\right]  ,
\end{equation}
where $\tilde{\Phi}$ is the massless gravidilaton field. Under the
substitution ${e^{-2\tilde{\Phi}}}=\varphi$, the above action reduces to the
JF Brans-Dicke action with the coupling parameter $\omega$ automatically fixed
to the value $\omega=-1$:
\begin{align}
S_{\text{JF}}  &  =\int{d^{4}x\sqrt{-\tilde{g}}\left[  {f(\varphi)}%
\widetilde{\mathbf{R}}+{h(\varphi)\tilde{g}^{\mu\nu}\varphi_{,\mu}%
\varphi_{,\nu}}\right]  }\nonumber\\
&  =\int{d^{4}x\sqrt{-\tilde{g}}\left[  {\varphi}\widetilde{\mathbf{R}}%
-{\frac{\omega}{\varphi}\tilde{g}^{\mu\nu}\varphi_{,\mu}\varphi_{,\nu}%
}\right]  },
\end{align}
where the gravitational coupling $f$ and the function $h$ have the forms as
above\footnote{An anonymous referee has raised an interesting question in
connection with reconciling the results obtained in this paper with those in
Ref.[19]. From the latter, it follows that no \textit{stable} wormhole
solution can exist in scalar-tensor theories with a ghost behaviour: Wormhole
solutions may exist, but with a negative gravitational coupling $f(\varphi
)=\varphi$ at least in some regions of the space leading to instabilities. The
solution discussed in the present paper does imply such ghost region since
$-2<\omega<-3/2$ corresponds in the Einstein frame to a scalar field with the
\textquotedblleft wrong\textquotedblright\ sign before the kinetic term.
Further, wormhole existence requires, by Ref.[19], that $f(\varphi)=\varphi
<0$, which can be ensured by assuming that the constant $\varphi_{0}$ in
Eq.(29) below be negative. Since there is no transition (because of the
exponential function there) from a positive to a negative value of the
gravitational coupling, the referee speculatively suggests that the wormhole
solution [Eqs.(26)-(30) below] could be stable. Intuitively, we do think that
the instability could appear due to such a transition and since there is no
transition in our solution, it may well represent a stable wormhole. The
authors of Ref.[19] consider that the instability appears due to a negative
pole of the \textit{effective potential }at the transition surface to
$f(\varphi)<0$. They state that this pole still does not guarantee
instability, and further studies are necessary. We note that $f(\varphi
)=\varphi$ is a Jordan frame variable in the action (2), and we have
\textit{not} considered any effective potential. Therefore, the answer is not
really immediate to us. On the other hand, in the Einstein frame, $\mu=-1$
(\textquotedblleft wrong sign\textquotedblright) is a necessary condition. In
this frame (and without any potential term), Gonzalez \textit{et al} [39]
analyzed the linear and nonlinear evolution of static, spherically symmetric
wormhole solutions for a massless ghost scalar field. They showed that all the
solutions are unstable with respect to linear and nonlinear spherically
symmetric perturbations and that the perturbation causes the wormholes to
either decay to a Schwarzschild black hole or undergo a rapid expansion.
However, only the zero mass wormhole could be stable, as argued by
Armend\'{a}riz-Pic\'{o}n [27]. The advantage with the Einstein frame is that
the equations are far simpler and hence convenient. Eventually, the results
concerning (in)stability in the Einstein frame have to be mapped back into the
Jordan frame to verify the extent to which those results hold true. This
requires a separate investigation.
\par
{}}. This particular value of $\omega$ is actually independent of the
dimensionality of the spacetime and the number of compactified dimensions
[25]\footnote{Lidsey \textit{et al} [25] define the Brans-Dicke scalar not as
$\varphi$ but as ${e^{-2\tilde{\Phi}}}$, showing that SF and JF are
equivalent, with the metric part remaining unaltered. They show that, more
generally, $\omega=-1+\frac{1}{d}$ where $d$ is the dimension of the
compactified torus on which gravidilation fields are zero. Also they argue
that $\omega$ is bounded by $-1\leq\omega<0$, where the lower bound
corresponds to the value that arises in the string effective action and is
formally saturated in the limit $d\rightarrow\infty$. From this viewpoint the
string value, $\omega=-1$, is a fixed-point under further dimensional
reduction by a finite number of dimensions.}. When $\widetilde{\Phi}=$
constant (frozen), all the frames (1), (2) and (5) coincide.

Let us unfix $\omega$ since we are not dealing with string theory here. Under
a further substitution
\begin{equation}
g_{\mu\nu}=\varphi\tilde{g}_{\mu\nu},
\end{equation}%
\begin{equation}
d\phi=\sqrt{\frac{2\omega+3}{2\mu}}\frac{{d\varphi}}{\varphi},\quad\quad
\alpha\neq0,
\end{equation}
the action (2) goes into the action in the Einstein frame (EF) or E-frame,
\begin{equation}
S_{\text{EF}}=\int{d^{4}x\sqrt{-g}}\left[  \mathbf{R}{-\mu g{}^{\mu\nu}%
\phi_{,\mu}\phi_{,\nu}}\right]  ,
\end{equation}
where we have introduced a constant arbitrary parameter $\mu$ ($=\pm1$) that
can have any sign. The field equations from action (5) are given by
\begin{equation}
R_{\mu\nu}=\mu\phi_{,\mu}\phi_{,\nu}%
\end{equation}%
\begin{equation}
\square^{2}\phi=0.
\end{equation}
We see that if the kinetic term ${\mu g{}^{\mu\nu}\phi_{,\mu}\phi_{,\nu}}$ in
the action has an overall positive sign, the stresses satisfy all energy
conditions and there are no wormholes. If however the kinetic term ${\mu
g{}^{\mu\nu}\phi_{,\mu}\phi_{,\nu}}$ has an overall negative sign, all the
energy conditions are violated giving rise to the possibility of wormhole
solutions. The situation is not as straightforward in the JF. Some energy
conditions can be locally violated in the JF even though there are no
violations (i.e., no ghost) in the EF [18,19].

The action (5) is also called the string action in the EF (see Gasperini and
Veneziano [26]). It seems remarkable that, starting from a general Lagrangian
and imposing only wormhole constraints (mainly the constraint of energy
violation), Armend\'{a}riz-Pic\'{o}n [27] derived the action (5) as the
simplest action for a general class of microscopic scalar fields. His
arguments have nothing to do with string theory, yet the end result is quite
the same. So we have here a hierarchy in which the physics of dilatonic
gravity meets that of wormholes.

From Eq.(4), it is obvious that the values of $\omega$ and $\mu$ can be chosen
independently. In the context of string theory, we must use only the model
independent, unique string value $\omega=-1$ [25]. In this case, we have
$d\phi=(1/\sqrt{2\mu})(d\varphi/\varphi)$, and the sign of $\mu$ is
essentially left undetermined by the string theory field equations in the EF,
viz., Eqs.(6) and (7). What actually determines the sign of $\mu$ is the
condition for energy violations by the $\phi-$ field stresses if we want
wormholes to exist. The violations are ensured by an overall negative sign
before the stress tensor, which can appear if $\phi$ is chosen as real
function and $\mu=-1$. Alternatively, if $\phi$ is chosen to be imaginary,
then $\mu=+1$. Either choice leads to a negative sign before the kinetic term
${\mu g{}^{\mu\nu}\phi_{,\mu}\phi_{,\nu}}$. However, no matter what the sign
of $\mu$ is, we can always move between the string actions phrased in SF and
EF respectively. Thus if there is a nonsingular wormhole in one, it can be
translated into the other frame.

As correctly concluded by Flanagan [28] using the most general framework,
classical observables are independent of the choice of different conformal
frames. But, since the vacuum JF action (2) (with any $\omega$) is essentially
incomplete because the matter part is absent, it has no conformal freedom due
to the fact that a conformal transformation does not keep track of the matter
part of the action. The vacuum theory itself may nevertheless be expressed in
different conformal frames\footnote{The class of conformal frames discussed by
Cho [29] and Faraoni [30] is a special case of the most general action
analyzed by Flanagan [28]. Faraoni [30] considered the conformal
transformations $\overline{g}_{\mu\nu}=\varphi^{2\xi}\widetilde{g}_{\mu\nu}$,
$\sigma=\varphi^{1-2\xi}$ under which the vacuum JF action (2) transfers to
the same form: $S_{\text{JF}}\rightarrow\overline{S}_{\text{JF}}=\int
d^{4}x(-\overline{g})^{1/2}\left[  \sigma\overline{\mathbf{R}}-\overline
{\omega}\sigma^{-1}\overline{g}^{\mu\nu}\sigma_{,\mu}\sigma_{,\nu}\right]  $
with $\overline{\omega}=\frac{\omega-6\xi(\xi-1)}{(1-2\xi)^{2}}$, where $\xi$
is an arbitrary real constant. Lidsey \textit{et al} [25] have provided a
higher dimensional generalization of this transfer, calling it conformal
invariance. Although there is a shift, ($\widetilde{g}_{\mu\nu}$, $\omega$,
$\varphi$) $\rightarrow$ ( $\overline{g}_{\mu\nu}$,$\overline{\omega}$,
$\sigma$) \textit{between} frames, the action itself is a scalar quantity so
that the physical content of the theory in any individual frame remains the
same. Note that Flanagan's general action is also form invariant [see her
\textit{Eqs. (2.2) }$\rightarrow$ \textit{(2.6)}] in exactly the same sense as
above, when all the relevant functions are suitably shifted. What essentially
remain conformal frame independent are the physical predictions, as stressed
by Lidsey \textit{et al }[25], Gasperini \& Veneziano [26] and Flanagan [28].
For instance, the string value $\omega=-1$ is a model independent prediction,
which remains the same in any individual conformal frame.} but, as pointed out
by Flanagan quoting Brans [31], only \textit{one} of those is a
\textquotedblleft physically correct frame\textquotedblright. \ So, which one
could it possibly be?

It is exactly in this context that the \textquotedblleft right
frame\textquotedblright\ proposed by Gasperini and Veneziano [26] becomes
relevant. In discussing pre- and post-big bang cosmology, they go from vacuum
string action (1) directly to (5) \textit{bypassing }the intermediate
$\omega=-1$ JF action (2) [\textit{their Eqs. (1.3) and (1.4)}] and conclude
that different classes of solutions do not correspond to different models of
pre-big bang inflation, but simply to different kinematical representations of
the same scenario in two different conformal frames. They show that different
frames are just related by a local (field dependent) conformal transformation
of the metric $-$ no physical observable should depend on it. This
notwithstanding, they show that there are important kinematical differences in
different frames. Inflation in the S-frame, for instance, can be represented
as gravitational collapse in the E-frame. In any case, they conclude that the
S-frame (whose metric coincides with the sigma model metric to which
fundamental strings are directly coupled) is the right frame\ because this is
the one offering the simplest intuitive picture of how things evolve and work.

Despite the logical gap between the argument of physical correctness and
simplicity, as far as frame independence of physical observables is concerned,
Flanagan's conclusion is well supported in the cosmological context with the
SF being favored on grounds of simplicity. We also take this standpoint in the
local wormhole gravity as well, only replacing SF by JF (the frames are
equivalent, as implied in [25], [26]). \ Note also that the 1962 list of exact
Brans' solutions have been proposed only in the vacuum JF and solar gravity
experiments have long been physically interpreted in this frame (see footnote
8). Since wormholes have not been ruled out, their interpretation in JF should
likewise be valid. In view of these arguments, we shall assume JF as the
single \textquotedblleft physically correct frame\textquotedblright\ in
the\textit{ vacuum }case since there is no conformal frame freedom here [28].
Certainly we are not stretching this assumption to conclude that JF is
synonymous with physical frame even when matter part is involved\footnote{We
have to point out that the question which frame, Einstein or Jordan, is more
physical has a long and controversial story and we do not take any position
here. Many authors treat Einstein frame as more of a mathematical
convenience.}.

We shall also show in the Appendix that the wormhole interpretation need not
essentially rely on the JF: The non-singular Class II wormhole in JF can be
mapped to a similar non-singular wormhole in the EF so that wormhole geometry
becomes conformal-frame-independent in these two frames. It would be nice to
have a general proof of this statement covering arbitrary conformal frames.
This concludes our discussion about conformal frames. Before we derive the
non-singular Brans wormhole out of Brans Class I solution, it will be useful
to provide an outline of what a wormhole is.

\textbf{III. Outline of wormhole geometry}

To be more informative, we define what a wormhole is, without going into a
full scale reporting. The notions stated in this brief review will be used
later. By definition, a wormhole is a topological short-cut tunnel connecting
two distant regions of a single spacetime or even two universes. The spacetime
has to satisfy certain constraints to qualify as a wormhole. We shall state
these constraints using the Morris and Thorne [32] canonical form for the
spacetime metric in \textquotedblleft standard\textquotedblright\ coordinates,
which is given by
\begin{equation}
d\tau^{2}=-e^{2\Phi(R)}dt^{2}+\left[  1-\frac{b(R)}{R}\right]  ^{-1}%
dR^{2}+R^{2}(d\theta^{2}+\sin^{2}\theta d\psi^{2}),
\end{equation}
where $\Phi(R)$ and $b(R)$ are redshift and shape functions, respectively, and
$R$ is defined by the positive circumferential radius $2\pi R$. The
terminology for $\Phi$ is self-evident. The space is spherical 3D at fixed
time $t$. So without loss of information, we concentrate on the $\theta=\pi
/2$, $t=$ constant 2D slice in which the metric reads%
\begin{equation}
d\tau^{2}=\left[  1-\frac{b(R)}{R}\right]  ^{-1}dR^{2}+R^{2}d\psi^{2}.
\end{equation}
We then remove the slice and embed it in the Euclidean 3D space having the
metric%
\begin{equation}
d\tau^{2}=\left[  1+\left(  \frac{dz}{dR}\right)  ^{2}\right]  dR^{2}%
+R^{2}d\psi^{2}.
\end{equation}
The isometry between (9) and (10) gives the shape of the axially symmetric
embedding surface $z=z(R)$ obtained by integrating%
\begin{equation}
\frac{dz}{dR}=\pm\left[  \frac{R}{b(R)}-1\right]  ^{-1/2}.
\end{equation}
The reason why $b(R)$ is called shape function is now clear. However, in most
natural situations and certainly in the Brans solutions, the integration does
not yield expressions $z=z(R)$ in a closed form. One would need to plot the
shape only by numerical calculation.

The basic constraints to be satisfied by a spacetime to qualify as a wormhole
are as follows:

\textit{(a) }The spacetime must have two asymptotically flat regions (mouths).

\textit{(b) }The mouths must be connected by a throat defined by the
\textit{minimum }circumferential radius. This occurs at a place $R=R_{0}$
where the vertical slope $\frac{dz}{dR}=\infty$. This implies that the throat
radius $R=R_{0}$ is a root of the equation $b(R_{0})=R_{0}$. The wormhole has
a hole of finite non-zero radius $R_{0}>0$, unlike a black hole center with
zero radius. Thus, $R\in\lbrack R_{0},+\infty)$.

\textit{(c)} The shape function must satisfy: $b(R)/R\rightarrow0$ as
$R\rightarrow\infty$. Also, $b(R)/R\leq1$ for all $R\geq$ $R_{0}$.

In order that the wormhole indeed flares out to two asymptotically flat space
times, the following condition%
\begin{equation}
\frac{d^{2}R}{dz^{2}}=\frac{b-b^{\prime}R}{2b^{2}}>0
\end{equation}
must be satisfied at or near the throat. This inequality imposes a constraint
on the type of source stress tensor $\mathbf{T}_{\mu\nu}$. Assuming an
isotropic scalar field stress tensor $\mathbf{T}_{\widehat{\mu}\widehat{\nu}%
}^{(\varphi)}=[\rho,p_{R},p_{\theta},p_{\psi}]$ in the static local
orthonormal frame ($\symbol{94}$) , the left hand side of vacuum Brans-Dicke
field equations $\mathbf{R}_{\widehat{\mu}\widehat{\nu}}-\frac{1}{2}%
\eta_{\widehat{\mu}\widehat{\nu}}\mathbf{R}=\mathbf{T}_{\widehat{\mu}%
\widehat{\nu}}^{(\varphi)}$ straightforwardly yield%
\begin{equation}
\rho=b^{\prime}/R^{2},p_{R}=[2(R-b)\Phi^{\prime}-b/R]/R^{2},p_{\theta}%
=p_{\psi}=p_{R}+(R/2)[(\rho+p_{R})\Phi^{\prime}+p_{R}^{\prime}],
\end{equation}
where primes denote differentiation relative to radius $R$, $\rho$ is the
scalar field energy density, $p_{R}$ is the radial tension, $p_{\theta
},p_{\psi}$ are transverse pressures, and $\eta_{\widehat{\mu}\widehat{\nu}}$
is the Minkowski metric in the orthonormal frame. \ Then the inequality in
(12) can be nicely rephrased by using the Morris and Thorne [32] function
$\zeta$ defined by $\zeta=-\frac{\rho+p_{R}}{\left\vert \rho\right\vert }$.
\ Putting in it the expressions from (13), combining it with the equality in
(12) and noting that $(R-b)\Phi^{\prime}\rightarrow0$ at $R=R_{0}$, we get at
the throat the following result
\begin{equation}
\zeta=\frac{2b^{2}}{R\left\vert b^{\prime}\right\vert }\frac{d^{2}R}{dz^{2}%
}=-\left(  \frac{\rho+p_{R}}{\left\vert \rho\right\vert }\right)  ,
\end{equation}
which shows that the flaring out condition (12) is satisfied only if
$\rho+p_{R}<0$. This violates a known energy condition since, for normal
matter, $\rho+p_{R}>0$ (null energy condition). A fast Lorentz boosted
traveller might see the violation as $\rho<0$, which means a violation of the
weak energy condition. Such energy condition violating matter is called
\textquotedblleft exotic\textquotedblright. Thus the flaring out constraint is:

\textit{(d)} We must have: $\rho<0$ and/or $\rho+p_{R}<0$ at least at or near
the throat.

\textit{(e)} There should be no horizon, that is, the redshift function $\Phi$
must be finite everywhere to prevent infinite redshift of signals from the
traveller to outside stationary observer.

\textit{(f) }The tidal forces (that are proportional to curvature tensor)
experienced by a traveller finite throughout the trip.

These are the main constraints to be satisfied if a given spacetime has to
represent a regular wormhole traversable in principle. Practical
traversability by humans requires further that tidal forces be tolerable, that
is, of the order of one Earth gravity.

\textbf{IV. Singular Brans I solution}

The field equations obtained by varying the vacuum JF action (2) are (dropping tilde)%

\begin{equation}
\mathbf{R}_{\mu\nu}-\frac{1}{2}g_{\mu\nu}\mathbf{R}=\frac{\omega}{\varphi^{2}%
}\left[  \varphi_{,\mu}\varphi_{,\nu}-\frac{1}{2}g_{\mu\nu}\varphi_{,\sigma
}\varphi^{,\sigma}\right]  +\frac{1}{\varphi}\left[  \varphi_{,\mu;\nu}%
-g_{\mu\nu}\square^{2}\varphi\right]  ,
\end{equation}%
\begin{equation}
\square^{2}\varphi=0.
\end{equation}
The general solution of these field equations, in isotropic coordinates
($t,r,\theta,\psi$), is taken in the form
\begin{equation}
d\tau^{2}=-e^{2\alpha(r)}dt^{2}+e^{2\beta(r)}[dr^{2}+r^{2}(d\theta^{2}%
+\sin^{2}\theta d\psi^{2})].
\end{equation}
The Brans class I solution [23] is given by%
\begin{equation}
e^{\alpha(r)}=e^{\alpha_{0}}\left[  \frac{1-B/r}{1+B/r}\right]  ^{\frac
{1}{\lambda}},
\end{equation}%
\begin{equation}
e^{\beta(r)}=e^{\beta_{0}}\left[  1+B/r\right]  ^{2}\left[  \frac
{1-B/r}{1+B/r}\right]  ^{\frac{\lambda-C-1}{\lambda}},
\end{equation}%
\begin{equation}
\varphi(r)=\varphi_{0}\left[  \frac{1-B/r}{1+B/r}\right]  ^{\frac{C}{\lambda}%
},
\end{equation}%
\begin{equation}
\lambda^{2}\equiv(C+1)^{2}-C\left(  1-\frac{\omega C}{2}\right)  >0,
\end{equation}
where $\lambda$, $\alpha_{0}$, $\beta_{0}$, $B$, $C$, and $\varphi_{0}$ are
real constants, and the radial marker $r\in(-\infty,+\infty$). The constants
$\alpha_{0}$ and $\beta_{0}$ are determined by asymptotic flatness at
$r=+\infty$ as $\alpha_{0}=$ $\beta_{0}=0$. The negative $r-$side needs a bit
of explanation. Note that the metric is invariant under inversion
$\overline{r}\rightarrow\frac{1}{r}$ for even values of the exponents
determined by $C(\omega)$, hence $\overline{r}=0$ is a second asymptotically
flat region. Defining a new coordinate chart there by $\overline{r}%
\rightarrow\frac{1}{\widehat{r}}$, which of course preserves the metric, we
find that $\widehat{r}\rightarrow-\infty$ \ represents the second
asymptotically flat region.\footnote{The following solution is the EF
counterpart of the singular Brans Class I solution (18)-(21): $d\tau
_{\text{EF}}^{2}=-\left(  \frac{1-\frac{m}{2r}}{1+\frac{m}{2r}}\right)
^{2\beta}dt^{2}+\left(  1-\frac{m}{2r}\right)  ^{2(1-\beta)}\left(  1+\frac
{m}{2r}\right)  ^{2(1+\beta)}[dr^{2}+r^{2}d\theta^{2}+r^{2}\sin^{2}\theta
d\psi^{2}],$ \ $\phi(r)=\sqrt{\frac{2(1-\beta^{2})}{\mu}}\ln\left[
\frac{1-\frac{m}{2r}}{1+\frac{m}{2r}}\right]  ,$ where $\mu$, $m$ and $\beta$
are arbitrary positive constants. This is known as Buchdahl solution [17],
rediscovered later as the singular Ellis I solution [22]. The metric is
invariant in form under inversion (for integer $\beta$) of the radial
coordinate $r\rightarrow\frac{m^{2}}{4r}$ and we have two asymptotically flat
regions (at $r=0$ and $r=\infty$), the minimum circumferential radius (throat)
occurring at $r_{0}=\frac{m}{2}\left[  \beta+\sqrt{\beta^{2}-1}\right]  $. The
reality of the wormhole throat is guaranteed by $\beta^{2}>1$, but this
condition also yields a naked singularity at $r=m/2$. For $\beta=1$, the
solution reduces to the Schwarzschild black hole. With regard to coordinate
patches, one must make a choice as to whether the isotropic coordinate lies in
the range $r\in$ ($m/2,\infty$) or $r\in$ ($0,m/2$). Though the patches are
disconnected, both separately are homeomorphic to the standard coordinate
$R\in$ ($2m,\infty$) and provide equivalent coverings, each with one
asymptotically flat end ($r\rightarrow0$ and $r\rightarrow+\infty$), of the
region $R>2m$ of the standard Buchdahl line element, variantly called
Janis-Newman-Winnicour line element. The same conclusions hold in the JF as
well, with $m/2=B^{\prime}$. We thank a referee for clarifying the coordinate
patches.} Unfortunately, the two asymptotically flat spacetimes on either side
are disconnected by the singularity at $r=\widehat{r}=B$, hence the solution
cannot be accepted as a wormhole, as mentioned before.

To see the naked singularity, it is enough to consider the Lorentz boost
invariant component of Riemann curvature in the freely falling orthonormal
frame ($\widehat{e}_{0^{\prime},}\widehat{e}_{1^{\prime},}\widehat
{e}_{2^{\prime},}\widehat{e}_{3^{\prime}}$), which turns out to be
\begin{equation}
\mathbf{R}_{\widehat{1}^{\prime}\widehat{0}^{\prime}\widehat{1}^{\prime
}\widehat{0}^{\prime}}=\frac{4Br^{3}Z^{2}[\lambda(r^{2}+B^{2})-Br(C+2)]}%
{\lambda^{2}(r^{2}-B^{2})^{4}},
\end{equation}
where%
\begin{equation}
Z\equiv\left(  \frac{r-B}{r+B}\right)  ^{(C+1)/\lambda}.
\end{equation}
Clearly, $\mathbf{R}_{\widehat{1}^{\prime}\widehat{0}^{\prime}\widehat
{1}^{\prime}\widehat{0}^{\prime}}\rightarrow\infty$ as $r\rightarrow B$. All
curvature invariants also exhibit this singular behavior. The Brans Class I
solution was already investigated in detail in [15], and criticized in [18],
so no need to repeat them here. We have still put the solution in view because
we would need it to generate a non-singular wormhole, which we do below.

\textbf{V. Non-singular Brans wormhole}

To obtain the non-singular solution, we need to first remove the above
mentioned singularity from the Class I solution. We can do it by the following
operations on it:%
\begin{equation}
r\rightarrow\frac{1}{r^{\prime}}\text{, }B\rightarrow\frac{i}{B^{\prime}%
}\text{, }\lambda\rightarrow-i\Lambda\text{, }\alpha_{0}\rightarrow\text{
}\epsilon_{0}\text{, }\beta_{0}\rightarrow\delta_{0}+2\ln B^{\prime},
\end{equation}
where $B^{\prime}$, $\Lambda$ are real. Using the identity%
\begin{equation}
\tan^{-1}(x)=\frac{i}{2}\ln\left(  \frac{1-ix}{1+ix}\right)  ,
\end{equation}
we arrive at the metric functions and the scalar field as follows%
\begin{equation}
d\tau^{2}=-e^{2\alpha(r^{\prime})}dt^{2}+e^{2\beta(r^{\prime})}[dr^{\prime
2}+r^{\prime2}(d\theta^{2}+\sin^{2}\theta d\psi^{2})],
\end{equation}
where
\begin{equation}
\alpha(r^{\prime})=\epsilon_{0}+\frac{2}{\Lambda}\tan^{-1}\left(
\frac{r^{\prime}}{B^{\prime}}\right)
\end{equation}%
\begin{equation}
\beta(r^{\prime})=\delta_{0}-\frac{2(C+1)}{\Lambda}\tan^{-1}\left(
\frac{r^{\prime}}{B^{\prime}}\right)  -\ln\left(  \frac{r^{\prime2}}%
{r^{\prime2}+B^{\prime2}}\right)
\end{equation}%
\begin{equation}
\varphi(r^{\prime})=\varphi_{0}\exp\left[  \frac{2C}{\Lambda}\tan^{-1}\left(
\frac{r^{\prime}}{B^{\prime}}\right)  \right]
\end{equation}%
\begin{equation}
\Lambda^{2}\equiv C\left(  1-\frac{\omega C}{2}\right)  -(C+1)^{2}>0.
\end{equation}
Asymptotic flatness at $r^{\prime}=\infty$ requires that
\begin{equation}
\epsilon_{0}=-\frac{\pi}{\Lambda}\text{, \ \ }\delta_{0}=\frac{\pi
(C+1)}{\Lambda}.
\end{equation}
The above solution has been listed by Brans [23] as his independent Class II
solution, but we see that the two classes are \textit{not} independent $-$ one
can be derived from the other by what we call Wick rotation (24). However,
though not independent, Class I and II solutions are by no means
\textit{equivalent} as the former is singular, while the latter is regular.

Next, it is necessary to ensure that the Wick rotation does not affect the
status of the new solution to be a solution of vacuum Brans-Dicke theory. This
can be seen from the following facts: The solution (18)-(20) satisfies the
field equations (15)-(16), \textit{only if} the constraint (21) connecting
various constants in the solution is satisfied. The transition $\lambda
\rightarrow-i\Lambda$ leads to just the Brans Class II solution for which new
constraint is (30), as can be found by directly putting the solution (27)-(29)
in the field equations or from the Brans' 1962 list itself. Therefore the
status of (27)-(29) as an exact solution of the field equations is confirmed.
It should be mentioned that the wormhole interpretation does not take away its
force to explain the observable predictions of usual stellar gravity. The
solution in the positive mass ($\frac{2B^{\prime}}{\Lambda}$) side does
explain the known weak field solar system predictions\footnote{To see this, it
is enough to consider the well known Eddington-Robertson expansion [33] of
\textit{any} metric produced by a static spherically symmetric body like the
Sun: $d\tau^{2}=-\left(  1-2\alpha\frac{M}{r^{\prime}}+2\beta\frac{M^{2}%
}{r^{\prime2}}+...\right)  dt^{2}+\left(  1-2\gamma\frac{M}{r^{\prime}%
}+...\right)  [dr^{\prime2}+r^{\prime2}(d\theta^{2}+\sin^{2}\theta d\psi
^{2})].$ Post-Newtonian Schwarzschild gravity corresponds to constants
$\alpha=\beta=\gamma=1$ . For the Brans Class II solution (27)-(29), the mass
$M=\frac{2B}{\Lambda}$ can be identified from the Newtonian limit
$e^{2\alpha(r^{\prime})}\simeq1-\frac{2M}{r^{\prime}}$. Under the same
approximation, we have $C=-\frac{1}{\omega+2}$, whence the Class II solution
leads to the values $\alpha=\beta=1$, $\gamma=\frac{\omega+1}{\omega+2}$.
Currently estimated value of $\gamma$ is $\gamma$ $=2\times(0.99992\pm
0.00014)-1$ [34], which is close to $1$ up to an accuracy of $10^{-4}$. The
purpose of saying all these is that the JF is a physically correct conformal
frame for the interpretation of observations in stellar gravity. As for
wormholes, they are not ruled out and the corresponding observations can
likewise be interpreted in JF. The important point is that the non-singular JF
Class II solution yields a similar non-singular solution in the EF (see
Appendix), but the latter expands on the positive mass side such that
$\alpha=\beta=\gamma=1$. These are exactly the values of post-Newtonian
Schwarzschild gravity with no question of Brans-Dicke coupling $\omega$
appearing anywhere. However, wormhole features still remain the same, as we
will show in the Appendix.}.

Under the radial coordinate transformation $r^{\prime}\rightarrow R$:%
\begin{equation}
R=r^{\prime}\exp[\beta(r^{\prime})]=r^{\prime}\left[  1+\frac{B^{\prime2}%
}{r^{\prime2}}\right]  \exp\left[  \delta_{0}-\frac{2(C+1)}{\Lambda}\tan
^{-1}\left(  \frac{r^{\prime}}{B^{\prime}}\right)  \right]  ,
\end{equation}
which shows that the circumferential radius $R\rightarrow\infty$\ as
$r^{\prime}\rightarrow\infty$, we obtain the redshift and shape functions
respectively as%
\begin{equation}
\Phi(R)=-\frac{\pi}{\Lambda}+\frac{2}{\Lambda}\tan^{-1}\left[  \frac
{r^{\prime}(R)}{B^{\prime}}\right]  ,
\end{equation}%
\begin{equation}
b(R)=R\left[  1-\left\{  1+\frac{2B^{\prime}}{r^{\prime2}(R)+B^{\prime2}%
}\left(  \frac{r^{\prime}(R)(C+1)}{\Lambda}-B^{\prime}\right)  \right\}
^{2}\right]  .
\end{equation}
Using these, we shall now examine all the wormhole constraints enumerated in
Sec. III.

\textbf{VI. Wormhole constraints}

(a) The solution set (27)-(29) is regular everywhere including at $r^{\prime
}=B^{\prime}$ as can be readily verified by computing the curvature
invariants. For instance, the Riemann curvature component invariant under
Lorentz boost is
\begin{equation}
\mathbf{R}_{\widehat{1}^{\prime}\widehat{0}^{\prime}\widehat{1}^{\prime
}\widehat{0}^{\prime}}=-\frac{4B^{\prime5}r^{\prime3}[\Lambda B^{\prime
2}+B^{\prime}r^{\prime}(C+2)-\Lambda r^{\prime2}]}{\Lambda^{2}(r^{\prime
2}+B^{\prime2})^{4}}\text{exp}\left[  \frac{4(C+1)}{\Lambda}\tan^{-1}\left(
\frac{r^{\prime}}{B^{\prime}}\right)  \right]  ,
\end{equation}
which is finite everywhere, and $\mathbf{R}_{\widehat{1}^{\prime}\widehat
{0}^{\prime}\widehat{1}^{\prime}\widehat{0}^{\prime}}\rightarrow0$ when
$r^{\prime}\rightarrow+\infty$ as well as when $r^{\prime}\rightarrow0$. This
second asymptotic region is evident also from the fact that $R=r^{\prime}%
$exp$(\beta r^{\prime})\rightarrow+\infty$ as\textit{ }$r^{\prime}%
\rightarrow0$.\footnote{We thank an anonymous referee for correcting an error
in the earlier version. This correction eventually led to a better
understanding of the coordinate cover on either side of the regular wormhole.
Essentially, it is a single coordinate cover but described in different
notations $r^{\prime}$($\rightarrow+\infty$) and $\widehat{r}$ ($\rightarrow
-\infty$) joined smoothly at $r^{\prime}=\widehat{r}=B^{\prime}$.
Alternatively, the asymptotic region $R\in\lbrack R_{0},+\infty)$ is covered
twice, once by $r^{\prime}\in\lbrack r_{0}^{\prime+},+\infty)$ and again
upside down by $r^{\prime}\in\lbrack0,r_{0}^{\prime+}]$. The homeomorphism
between the two covers is evident.} All the curvature invariants are also
finite and go to zero in these limits. The latter limit implies that
$r^{\prime}=0$ is the second asymptotically flat region. \ This region is seen
in the metric under the transformation $r^{\prime}\rightarrow1/\widehat{r}$
using the identities%
\begin{align}
\tan^{-1}\left(  \frac{1}{x}\right)   &  \equiv\frac{\pi}{2}-\tan
^{-1}(x)\text{, \ \ \ \ \ \ \ \ for }x>0\\
\tan^{-1}\left(  \frac{1}{x}\right)   &  \equiv-\frac{\pi}{2}-\tan
^{-1}(x)\text{, \ \ \ \ \ \ for }x<0,
\end{align}
and then taking the limit $\widehat{r}\rightarrow-\infty$. \ Hence, the
solution is twice asympotically flat satisfying the constraint \textit{(a)}.

(b) The throat occurs at the coordinate marker\footnote{We sincerely thank a
referee for pointing out an error in the earlier version. There would be a
negative sign before the first term in the bracket of Eq.(38), which was
missing in the earlier version. However, there is no alteration in the
conclusions.}%
\begin{equation}
r_{0}^{\prime\pm}=B^{\prime}\left[  -\frac{C+1}{\Lambda}\pm\sqrt{1+\left(
\frac{C+1}{\Lambda}\right)  ^{2}}\right]  ,\text{ \ }B^{\prime}>0
\end{equation}
obtained by the minimizing the circumferential radius $R=r^{\prime}\exp
[\beta(r^{\prime})]$. We shall discard the values $r_{0}^{\prime-}$ because it
is always negative thereby leading to unphysical negative circumferential
radius $2\pi R_{0}^{-}$. The throat $r_{0}^{\prime+}$ is always positive
because $\sqrt{1+\left(  \frac{C+1}{\Lambda}\right)  ^{2}}>\frac{C+1}{\Lambda
}$ and thus can take on any positive nonzero value leading to\footnote{The
radial coordinate $r^{\prime}$ is an abstract coordinate chart covering the
entire space, whereas the Morris-Thorne radius $R$ is defined by physically
measurable circumference but it does not cover the entire space. The throat
radius can be calculated either by the minimal circumferential radius $2\pi R$
or from the shape function $b(R)$. Both of course yield the same answer.}
\begin{equation}
R_{0}^{+}=r_{0}^{\prime+}\left[  1+\frac{B^{\prime2}}{r_{0}^{\prime+2}%
}\right]  \exp\left[  \delta_{0}-\frac{2(C+1)}{\Lambda}\tan^{-1}\left(
\frac{r_{0}^{\prime+}}{B^{\prime}}\right)  \right]  >0.
\end{equation}
It can be verified that the weak field expansions of $e^{2\alpha(r^{\prime})}$
using the identities (35), (36) yield respectively the asymptotic
Schwarzschild masses $M^{\pm}$
\begin{align}
M^{+}  &  =\frac{2B^{\prime}}{\Lambda}\text{ \ as }r^{\prime}\rightarrow
+\infty\text{,}\\
M^{-}  &  =-\frac{2B^{\prime}}{\Lambda}\text{exp}\left[  \frac{2\pi}{\Lambda
}\right]  \text{ \ as }\widehat{r}\rightarrow-\infty
\end{align}
of two \textit{asymmetric }mouths lying on either side of the throat
$r=r_{0}^{\prime+}$. Hence the contraint\textit{ (b)} is satisfied.

(c) It can be directly seen that the shape function satisfies:
$b(R)/R\rightarrow0$ as $R\rightarrow\infty$ and $b(R)/R\leq1$ for all $R\geq$
$R_{0}$. Hence \textit{(c) }is satisfied.

(d) The stress components (13) can be computed from (27)-(29) giving the
scalar field energy density and radial pressure as%
\begin{equation}
\rho=-\frac{4B^{\prime6}r^{\prime4}[(C+1)^{2}+\Lambda^{2}]}{\Lambda
^{2}(r^{\prime2}+B^{\prime2})^{4}}\text{exp}\left[  \frac{4(C+1)}{\Lambda}%
\tan^{-1}\left(  \frac{r^{\prime}}{B^{\prime}}\right)  \right]  ,
\end{equation}%
\begin{align}
\rho+p_{R}  &  =-\frac{4B^{\prime5}r^{\prime3}[\Lambda C(r^{\prime2}%
-B^{\prime2})+2B^{\prime}r^{\prime}(C+1+\Lambda^{2})]}{\Lambda^{2}(r^{\prime
2}+B^{\prime2})^{4}}\nonumber\\
&  \times\text{exp}\left[  \frac{4(C+1)}{\Lambda}\tan^{-1}\left(
\frac{r^{\prime}}{B^{\prime}}\right)  \right]  .
\end{align}
It is evident that $\rho<0$ everywhere in the spacetime and so weak energy
condition is violated. Hence \textit{(d)} is satisfied, whatever be the value
of a real $\Lambda$.

Now, the weak energy condition ($\rho>0$) is stronger than the null energy
condition ($\rho+p_{R}>0$) and the violation of the latter is proven to be
also the minimal violation required of a wormhole [35]. To ensure that
$\rho+p_{R}<0$ at or near the throat, we would need to fix the values of
$\Lambda,C$ and $r_{0}^{\prime+}$ in terms of $\omega$. To this end, using the
weak field value
\begin{equation}
C=-\frac{1}{\omega+2},
\end{equation}
we get%
\begin{equation}
\Lambda=\pm\sqrt{-\frac{2\omega+3}{2\omega+4}}.
\end{equation}
The condition that $\Lambda$ be real immediately yields a new range
$-2<\omega<-3/2$. Outside this range, $\Lambda$ becomes imaginary, which in
turn contradicts the Wick rotation (24). Hence those values outside the above
interval for $\omega$ do not correspond to wormholes. Putting the values of
$C$ and $\Lambda$, together with any value of $\omega$ in the allowed range,
we would obtain two values for the throat $r_{0}^{\prime\pm}$, one positive
and the other negative, for real $B^{\prime}$. The negative value is of course
discarded. Figs.1 and 2 show that both $\rho<0$ and $\rho+p_{R}<0$ in the new
range $-2<\omega<-3/2$. Both $\rho$ and $\rho+p_{R}$ go to zero at the
asymptotic ends $r^{\prime}\rightarrow+\infty$ as well as at $\widehat
{r}\rightarrow-\infty$, as expected.

(e) It is evident from $\Phi$ in (33) that it is always finite everywhere
including at the asymptotic limits $r^{\prime}=0$ and $\infty$. \ Hence
\textit{(e)} is satisfied.

(f) The differential of the radial tidal acceleration $\Delta a^{r}$ in the
static orthonormal frame ($\widehat{e}_{t,}\widehat{e}_{R,}\widehat{e}%
_{\theta,}\widehat{e}_{\varphi}$) is given by%
\begin{equation}
\Delta a^{r}=-\mathbf{R}_{\widehat{R}\widehat{t}\widehat{R}\widehat{t}}\xi
^{R},
\end{equation}
where $\xi^{R}$ is the radial component of the separation vector and the
curvature tensor component is given by [32]%
\begin{equation}
\mathbf{R}_{\widehat{R}\widehat{t}\widehat{R}\widehat{t}}=(1-b/R)\left\{
-\Phi^{\prime\prime}+\frac{b^{\prime}R-b}{2R(R-b)}\Phi^{\prime}-(\Phi^{\prime
})^{2}\right\}  .
\end{equation}
This component is invariant under a Lorentz boost [16,32], $\mathbf{R}%
_{\widehat{1}^{\prime}\widehat{0}^{\prime}\widehat{1}^{\prime}\widehat
{0}^{\prime}}=\mathbf{R}_{\widehat{R}\widehat{t}\widehat{R}\widehat{t}}$. For
the metric given by (27)-(29), we already calculated in (35) the radial tidal
acceleration $\mathbf{R}_{\widehat{1}^{\prime}\widehat{0}^{\prime}\widehat
{1}^{\prime}\widehat{0}^{\prime}}$ and showed that it is always finite in the
Lorentz boosted orthonormal frame ($\widehat{e}_{0^{\prime},}\widehat
{e}_{1^{\prime},}\widehat{e}_{2^{\prime},}\widehat{e}_{3^{\prime}}$) of the traveller.

The lateral tidal forces in the Lorentz boosted frame are [16,32]
\begin{align}
\mathbf{R}_{\widehat{2}^{\prime}\widehat{0}^{\prime}\widehat{2}^{\prime
}\widehat{0}^{\prime}}  &  =\mathbf{R}_{\widehat{\theta}\widehat{t}%
\widehat{\theta}\widehat{t}}+\left(  \frac{v^{2}}{1-v^{2}}\right)
(\mathbf{R}_{\widehat{\theta}\widehat{t}\widehat{\theta}\widehat{t}%
}+\mathbf{R}_{\widehat{\theta}\widehat{R}\widehat{\theta}\widehat{R}})\\
\mathbf{R}_{\widehat{3}^{\prime}\widehat{0}^{\prime}\widehat{3}^{\prime
}\widehat{0}^{\prime}}  &  =\mathbf{R}_{\widehat{\varphi}\widehat{t}%
\widehat{\varphi}\widehat{t}}+\left(  \frac{v^{2}}{1-v^{2}}\right)
(\mathbf{R}_{\widehat{\varphi}\widehat{t}\widehat{\varphi}\widehat{t}%
}+\mathbf{R}_{\widehat{\varphi}\widehat{R}\widehat{\varphi}\widehat{R}}),
\end{align}
where $v$ is the instantaneous velocity of the traveller. Since, by spherical
symmetry, $\mathbf{R}_{\widehat{\theta}\widehat{t}\widehat{\theta}\widehat{t}%
}=\mathbf{R}_{\widehat{\varphi}\widehat{t}\widehat{\varphi}\widehat{t}}$ and
$\mathbf{R}_{\widehat{\theta}\widehat{R}\widehat{\theta}\widehat{R}%
}=\mathbf{R}_{\widehat{\varphi}\widehat{R}\widehat{\varphi}\widehat{R}}$, we
get, for the canonical metric (8):
\begin{equation}
\mathbf{R}_{\widehat{2}^{\prime}\widehat{0}^{\prime}\widehat{2}^{\prime
}\widehat{0}^{\prime}}=\mathbf{R}_{\widehat{3}^{\prime}\widehat{0}^{\prime
}\widehat{3}^{\prime}\widehat{0}^{\prime}}=\frac{1}{2R^{2}(1-v^{2})}\left[
v^{2}\left(  b^{\prime}-\frac{b}{R}\right)  +2(R-b)\Phi^{\prime}\right]  .
\end{equation}
For $v<1$, they remain finite. Since all the tidal forces are finite for a
trip across, the constraint \textit{(f)} is satisfied.

Thus the Brans Class II solution represents a regular wormhole, traversable
\textit{in principle}. However, the requirement of comfortable trip by humans
\textit{in practice}, viz., that the tidal forces be tolerable (not exceeding
one Earth gravity) is a separate question. We shall discuss the relevant
magnitudes in the context of analogue wormhole.

\textbf{VII. Wormhole analogy}

First, let us state what we mean by naked black holes. The idea was first
discussed by Horowitz and Ross [16]. They defined the naked black hole as a
spacetime in which an infalling observer meets the maximum tidal force not at
the horizon but \textit{above} it. In a freely falling frame, the curvature
components could be larger than those at the horizon. Normally, tidal forces
are maximum at the classically invisible horizon. Since the region of large
tidal forces is visible to distant observers, Horowitz and Ross called such
objects \textquotedblleft naked black holes\textquotedblright. In our case,
the role of invisible horizon is played by the visible throat of the analogue wormhole.

Next, to see the wormhole analogy, let us assess the variation in the
curvature components above the throat by explicitly calculating the components
in the infalling orthonormal frame ($\widehat{e}_{0^{\prime},}\widehat
{e}_{1^{\prime},}\widehat{e}_{2^{\prime},}\widehat{e}_{3^{\prime}}$). The
results are:%

\begin{equation}
\mathbf{R}_{\widehat{1}^{\prime}\widehat{0}^{\prime}\widehat{1}^{\prime
}\widehat{0}^{\prime}}=-\frac{4B^{\prime5}r^{\prime3}[\Lambda B^{\prime
2}+B^{\prime}r^{\prime}(C+2)-\Lambda r^{\prime2}]}{\Lambda^{2}(r^{\prime
2}+B^{\prime2})^{4}}\text{exp}\left[  \frac{4(C+1)}{\Lambda}\tan^{-1}\left(
\frac{r^{\prime}}{B^{\prime}}\right)  \right]  , \tag{35}%
\end{equation}%
\begin{align}
\mathbf{R}_{\widehat{2}^{\prime}\widehat{0}^{\prime}\widehat{2}^{\prime
}\widehat{0}^{\prime}}  &  =\mathbf{R}_{\widehat{3}^{\prime}\widehat
{0}^{\prime}\widehat{3}^{\prime}\widehat{0}^{\prime}}=\frac{2B^{\prime
5}r^{\prime3}\text{exp}\left[  \frac{4(C+1)}{\Lambda}\tan^{-1}\left(
\frac{r^{\prime}}{B^{\prime}}\right)  \right]  }{\Lambda^{2}(r^{\prime
2}+B^{\prime2})^{4}(1-v^{2})}\times\nonumber\\
&  [\Lambda(r^{\prime2}-B^{\prime2})\{(C+1)v^{2}-1\}+2B^{\prime}r^{\prime
}\{C+1+\Lambda^{2}v^{2}\}].
\end{align}

Putting the value of $C$ and either value of $\Lambda$ in turn we can express
$\left\vert \mathbf{R}_{\widehat{1}^{\prime}\widehat{0}^{\prime}\widehat
{1}^{\prime}\widehat{0}^{\prime}}\right\vert =\left\vert g\mathbf{(}%
\omega,r^{\prime},B^{\prime})\right\vert $, where the function $g$ results
from the right hand side of (35). The behavior of $g$ in Fig.3 then exhibits
the wormhole analogue of the naked black hole. For positive $\Lambda$,
curvature increases above the throat resembling the curvature enhancement
above the horizon in naked black holes. For negative $\Lambda$, there is no
enhancement above the throat, as measured by the infalling observer, and hence
no such analogy is possible (Fig.4).

Regardless of the implications of wormhole analogy, the regular Brans II
wormhole is traversable only in principle, not in practice. To be suitable for
comfortable travel by a human of length $2$ meters, the radial tidal
acceleration should be roughly of the order of one Earth gravity $g_{\oplus}$
($=\frac{GM_{\oplus}}{r_{\oplus}^{2}}\simeq980$ cm/sec$^{2}$). In relativistic
units, it should be $\frac{g_{\oplus}}{c^{2}}$. Since $\xi^{R}$ has the
dimension $L$ of length, looking at (46), we get that the dimension of
$\mathbf{R}_{\widehat{1}^{\prime}\widehat{0}^{\prime}\widehat{1}^{\prime
}\widehat{0}^{\prime}}$ must be $\frac{g_{\oplus}}{c^{2}\times L}\sim L^{-2}$.
Therefore, the right hand side Eq.(28) should be less than $\lesssim
\frac{g_{\oplus}}{c^{2}\times2\text{ mtrs}}\sim10^{-20}$cm$^{-2}$ [32], which
requires that the magnitude of the curvature components should be very close
to zero in the orthonormal frame of the traveller. Such a condition is easily
provided by $\Phi=0$, which is not the case here. To have an idea of the
magnitudes involved, we focus on a typical value in Fig.3, say, $\omega=-1.7$
with positive $\Lambda$ and observe that $\left\vert g\right\vert =$ $0.001$
cm$^{-2}$ at $r_{0}^{+}=0.17$. This is $10^{17}$ orders of magnitude more than
the value is required ($10^{-20}$cm$^{-2}$) for practical travel, which is too
large to be comfortable with! The maximum of $\left\vert g\right\vert $ occurs
up nearby, at $r^{\prime}=0.2$, where $\left\vert g\right\vert =0.002$
cm$^{-2}$. Though insignificant, it's still an enhancement in radial curvature
over the value at $r_{0}^{+}$, supporting the analogy in principle. However,
no such enhancement is observed when $\Lambda$ is negative, as shown in Fig.4.

We can see how the lateral forces fix the velocity $v$ of an inanimate
particle especially at the highest curvature at $r^{\prime}=0.2$. Assuming
that the radial and lateral accelerations are of the same order, we have the
condition%
\begin{equation}
\left\vert \mathbf{R}_{\widehat{2}^{\prime}\widehat{0}^{\prime}\widehat
{2}^{\prime}\widehat{0}^{\prime}}\right\vert \leq0.002.
\end{equation}
Putting the value $\omega=-1.7$ and in units $B^{\prime}=1$, we get from (50)%
\begin{equation}
0.0032-0.0044v^{2}\leq0.002(1-v^{2}),
\end{equation}
which yields $0.59\leq v<1$.

\textbf{VIII. Summary}

There has been a controversy over several years as to which frame, JF or EF,
is physical. While the aim of the present paper is not to conclude this
debate, we think that physical predictions should in general be
conformal-frame-independent where conformal freedom is allowed. We chose JF to
work out the wormhole solution for two reasons: On the one hand, the vacuum
case has no conformal freedom, which means that only one frame has to be
physical [28], and on the other, weak field solar gravity predictions have
been well interpreted in the JF [34]. In short, we adhered to the Brans-Dicke
theory in the frame proposed originally by its progenitors on the basis of
Machian philosophy. In the EF, the theory is no longer Machian, the parameter
$\omega$ disappears leading indistinguishably to the post-Newtonian
Schwarzschild values. Regardless of this, the same wormhole geometry is well
preserved also in the EF (see Appendix). \ Therefore, the validity of our
analysis need not be treated as restricted only to JF.

Now we come to the main thrust of the paper. From Sec.III downwards, we have
systematically developed what constraints are to be satisfied by a spacetime
to qualify as a regular wormhole. Several related important questions are
addressed in the footnotes in order that a reader can smoothly follow the
string of developments in the text. Thus, after reviewing the basic wormhole
constraints, we have derived a regular wormhole solution from the singular
Brans Class I solution. We find that the new regular solution is just the
Brans Class II solution to be found in his list of independent solutions of
the vacuum theory. However, since the passage Class I $\leftrightarrow$ Class
II is possible via Wick rotations, one might not regard them as strictly
independent solutions. This is a new information, probably unknown at large.

We have demonstrated that the regular Class II solution nicely satisfies all
the constraints required of a wormhole, while preserving the observed weak
field post-Newtonian values in the positive mass mouth. We see that $\rho<0$
everywhere [see Eq.(42)], no matter what the value of $\omega$ is. The
bottomline is that the solution is a born wormhole for \textit{any} $\omega$
as long as we don't specify $C(\omega)$ and remain happy with the violation of
weak energy condition alone. But specify $C(\omega)$ we must if we want a
tally with, say, solar system observations. In this case, limits on $\omega$
will appear. Accordingly, we chose the weak field equation (44), and the
requirement that $\Lambda(\omega)$ be real immediately yields the range
$-2<\omega<-3/2$. We observe an interesting thing happening in this range: The
minimal violation of energy condition, viz., $\rho+p_{R}<0$, though not
mandatory, is respected in addition to $\rho<0$ (see Figs.1 \& 2). It is
however quite likely that the $\omega-$ range would be different for different
$C(\omega)$ functions (see Ref.[10] for details on this point). Such
specifications of $C(\omega)$ seem always possible under different physical
circumstances, see [38].

So, all in all, we have a true Brans-Dicke wormhole with a stellar sized mouth
that can act as a launching pad for passage through the throat out into the
negative mass mouth. The wormhole is like a Janus-faced object that attracts
matter at one end and spews out at the other. A hypothetical traveller has to
accelerate to get out of the spacetime of attractive positive mass ($M^{+}$),
and after passing the throat, decelerate to come to a stop in the spacetime of
repulsive negative mass ($M^{-}$). We think that the Class II solution is a
remarkable example of a \textit{natural} wormhole (as distinct from
artificially constructed ones, such as those in Morris and Thorne [32])
provided by the Brans-Dicke theory. If the theory is true, the gravity-scalar
field coupling via $\omega$ does lead to the possibility of exotic matter
being present in the spacetime. However, since $\Phi\neq0$, human travel seems
impossible as the tidal forces would be intolerable at or near the throat.
Only an inanimate sufficiently hard test object (say, a robot) can pass
through to the other side.

Finally, we have shown that there can be a wormhole analogy to Horowitz-Ross
naked black holes when $\Lambda>0$ (Fig.3). But this analogy is more academic
than practical because the rise in curvature compared to that in the throat is
too tiny. No such rise is observed when $\Lambda<0$ (Fig.4). Several questions
still need to be answered. For instance, the possibility of naked black holes
can impact the problem of information puzzle at the horizon, as speculated by
those authors. What will be its parallel with wormholes, remembering that
wormholes are not collapsed objects? How to have a significant enhancement in
curvature? Will a different $C(\omega$) valid near strong gravity [38] such as
that of a neutron star do? These we reserve as tasks for the future.

\textbf{Acknowledgments}

We sincerely thank Professor A.A. Starobinsky for several informative
correspondences. One of us (AB) thanks the CSIR, Government of India for
financial support through NET-JRF fellowship. KKN acknowledges the financial
support provided by the GNFM-INDAM grant of the Government of Italy under
which major part of the work has been carried out at the University of
Salerno. All authors are deeply indebted to Guzel N. Kutdusova and Sonali
Sarkar for their moral and technical support.

\textbf{Figure captions}

Fig.1. We take unit as $B^{\prime}=1$, the positive sign before $\Lambda$ and
a typical value in the new range $-2<\omega<-3/2$, say, $\omega=-1.7$. The
positive value of the throat radius is $r_{0}^{\prime+}=0.17$, $C=-3.33$,
$\Lambda=0.81$ and $R_{0}^{+}=2.31$. We see that both $\rho<0$ and $\rho
+p_{R}<0$ around the throat.

Fig.2. We take unit as $B^{\prime}=1$, the negative sign before $\Lambda$ and
a typical value in the new range $-2<\omega<-3/2$, say, $\omega=-1.7$. The
positive value of the throat radius is $r_{0}^{\prime+}=5.88$, $C=-3.33$,
$\Lambda=-0.81$ and $R_{0}^{+}=15.84$. We see that both $\rho<0$ and
$\rho+p_{R}<0$ everywhere.

Fig.3. We take unit as $B^{\prime}=1$, the positive sign before $\Lambda$ and
a typical value in the new range $-2<\omega<-3/2$, say, $\omega=-1.7$. The
positive value of the throat radius is $r_{0}^{\prime+}=0.17$, $C=-3.33$ and
$\Lambda=0.81$. The maximum value of $\mathbf{R}_{\widehat{1}^{\prime}%
\widehat{0}^{\prime}\widehat{1}^{\prime}\widehat{0}^{\prime}}$ occurs at
$r^{\prime}=0.2$, which lies above the throat $r_{0}^{\prime+}$. The plot
shows curvature enhancement away from from the throat.

Fig.4. We take unit as $B^{\prime}=1$, the negative sign before $\Lambda$ and
a typical value in the new range $-2<\omega<-3/2$, say, $\omega=-1.7$. The
positive value of the throat radius is $r_{0}^{\prime+}=5.88$, $C=-3.33$ and
$\Lambda=-0.81$. The plot shows a steady decrease in curvature $\left\vert
\mathbf{R}_{\widehat{1}^{\prime}\widehat{0}^{\prime}\widehat{1}^{\prime
}\widehat{0}^{\prime}}\right\vert $ away from the throat, hence no analogy
exists in this case.

\begin{center}
\textbf{Appendix}

\textbf{The EF variant of the JF Brans II solution}
\end{center}

In the main text, the Class II solution was interpretated as a regular
wormhole only in the JF. But this is no limitation. As an illustration, we may
go over to the EF to see that the same interpretation still holds as detailed below.

After redefining the constants
\begin{equation}
\frac{2C}{\Lambda}\rightarrow4\lambda_{1},\frac{2(C+2)}{\Lambda}%
\rightarrow4\gamma_{1},-\frac{2\pi}{\Lambda}\rightarrow\epsilon_{1},\frac
{2\pi(1+C)}{\Lambda}\rightarrow\zeta_{1},
\end{equation}
the JF solution (27)-(29) reads in the conformally rescaled EF as follows:%
\begin{equation}
d\tau_{\text{EF}}^{2}=-P(r^{\prime})dt^{2}+Q(r^{\prime})[dr^{\prime
2}+r^{\prime2}(d\theta^{2}+\sin^{2}\theta d\psi^{2})],
\end{equation}
where%
\begin{align}
P(r^{\prime})  &  =\exp\left[  2\epsilon_{1}+4\gamma_{1}\tan^{-1}(r^{\prime
}/B^{\prime})\right]  ,\text{ }\\
Q(r^{\prime})  &  =\left(  1+\frac{B^{\prime2}}{r^{\prime2}}\right)  ^{2}%
\exp\left[  2\zeta_{1}-4\gamma_{1}\tan^{-1}(r^{\prime}/B^{\prime})\right]  ,\\
\phi(r^{\prime})  &  =4\lambda_{1}\tan^{-1}(r^{\prime}/B^{\prime})\text{,
\ where }2\lambda_{1}^{2}=1+\gamma_{1}^{2}\text{. }%
\end{align}
Asymptotic flatness requires that $\epsilon_{1}=-\pi\gamma_{1}$ and $\zeta
_{1}=\pi\gamma_{1}$. The constraint equation $2\lambda_{1}^{2}=1+\gamma
_{1}^{2}$ among free constants comes from the EF field equations when the
solution is put into them. This is the EF counterpart of the JF\ Class II
solution. To make it look more familiar, \ transform the radial variable as
$\ell=r^{\prime}+\frac{B^{\prime2}}{r^{\prime}}$. Then the solution (56)-(58)
goes over into%
\begin{align}
d\tau_{\text{EF}}^{2}  &  =-F(\ell)dt^{2}+F^{-1}(\ell)[d\ell^{2}+(\ell
^{2}+m^{2})(d\theta^{2}+\sin^{2}\theta d\psi^{2})],\text{ }\\
F(\ell)  &  =\exp\left[  -2\pi\gamma_{1}+4\gamma_{1}\tan^{-1}\left(
\frac{\ell+\sqrt{\ell^{2}+m^{2}}}{m}\right)  \right]  ,\text{ }\\
\phi(\ell)  &  =4\lambda_{1}\tan^{-1}\left(  \frac{\ell+\sqrt{\ell^{2}+m^{2}}%
}{m}\right)  ,
\end{align}
where we have identified $m=2B^{\prime}$ and $\ell\in(-\infty,+\infty)$.
Following the same steps as in Sec.V, it can be straightaway verified that
\textit{this is a twice asymptotically flat regular wormhole similar to that
in the JF}. So the features of wormhole geometry are conformally well
preserved, as we had promised to show in Sec.II. When $m=0$, the spacetime is
flat, as expected. When $m\neq0$ but $\gamma_{1}=0$, we obtain what is known
as a \textquotedblleft zero total mass wormhole\textquotedblright, in which
the pure scalar field masses of opposite signs at either side add exactly to
zero. Its lensing properties have been studied in Refs. [36,37].

The Eqs.(60), (61) are \textit{exactly} the same as the regular
Ellis-Bronnikov wormhole [22] in the EF already given by%
\begin{align}
F_{\text{Ellis}}(\ell) &  =\exp\left[  -2\pi\gamma_{1}+4\gamma_{1}\tan
^{-1}\left(  \frac{\ell}{m}\right)  \right]  ,\text{ }\\
\phi_{\text{Ellis}}(\ell) &  =4\lambda_{1}\tan^{-1}\left(  \frac{\ell}%
{m}\right)  \text{, }%
\end{align}
both the sets identically satisfying the EF field equations (6), (7):
\begin{align}
\left[  \left(  \ell^{2}+m^{2}\right)  \frac{F^{\prime}}{F}\right]  ^{\prime}
&  =0,\\
\left(  \frac{F^{\prime}}{F}\right)  ^{2}+\frac{4m^{2}}{\left(  \ell^{2}%
+m^{2}\right)  ^{2}}-2\phi^{2} &  =0,
\end{align}
yielding the same constraint equation $2\lambda_{1}^{2}=1+\gamma_{1}^{2}$,
where primes denote differentiation with respect to $\ell$. The zero mass case
is also the same. The point we want to clarify is that the regular wormhole
derived in this article need not rely exclusively on the JF but, as we have
shown, the EF conformal variant yields exactly the same regular wormhole as
well. The identity of solutions (60), (61) with those in (62), (63) can be
easily seen by redefintion of variables and arctan identities.

However, note that the two sets of solutions have been obtained from the same
JF\ Class I but by two different routes:

JF\ Class I $\overset{\text{conf. trans.}}{\rightarrow}$ Buchdahl
$\overset{\text{coord. trans.+Wick}}{\rightarrow}$ Ellis-Bronnikov wormhole
(62), (63) in EF (Ref.[6]).

JF\ Class I $\overset{\text{Wick rotation}}{\rightarrow}$ JF\ Class II
$\overset{\text{conf.trans.+coord.trans.}}{\rightarrow}$New solution (60)-(61)
in EF (this paper).

The identity of the end result shows that the operations are commutative.

\textbf{References}

[1] A.G. Agnese and M. La Camera, Phys. Rev. D \textbf{51,} 2011 (1995).

[2] M. Visser and D. Hochberg, \textit{Proc. Haifa Workshop on the Internal
Structure of Black Holes and Space Time Singularities }(Jerusalem, Israel,
June, 1997) [arXiv:gr-qc/970001], p.20.

[3] K.K. Nandi, A. Islam, and J. Evans, Phys. Rev. D \textbf{55}, 2497 (1997).

[4] K.K. Nandi, B. Bhattacharjee, S.M.K. Alam, and J. Evans, Phys. Rev. D
\textbf{57}, 823 (1998).

[5] A. Bhadra, K. Sarkar, D. P. Datta, and K. K. Nandi, Mod. Phys. Lett. A
\textbf{22} (2007) 367 [arXiv:gr-qc/0605109].

[6] Kamal K. Nandi, Ilnur Nigmatzyanov, Ramil Izmailov and Nail G. Migranov,
Class.Quant.Grav. \textbf{25},165020 (2008).

[7] Ernesto F. Eiroa, Martin G. Richarte, and Claudio Simeone, Phys. Lett. A
\textbf{373}, 1 (2008); \textit{ibid.} A \textbf{373} E2399 (2009).

[8] Arunava Bhadra, Ion Simaciu, Kamal Kanti Nandi, and Yuan-Zhong Zhang,
Phys. Rev. D \textbf{71}, 128501 (2005).

[9] L.A. Anchordoqui, S. P. Bergliaffa, and D.F. Torres, Phys. Rev. D
\textbf{55,} 526 (1997).

[10] Francisco S. N. Lobo and Miguel A. Oliveira, Phys. Rev. D \textbf{81},
067501 (2010).

[11] J.P.S. Lemos, F.S.N. Lobo, and S.Q. de Oliveira, Phys. Rev. D \textbf{68}
064004 (2003).

[12] F.S. N. Lobo, Phys. Rev. D \textbf{71}, 084011 (2005); Phys. Rev. D
\textbf{73}, 064028 (2006).

[13] S. V. Sushkov, Phys.Rev. D \textbf{71}, 043520 (2005).

[14] Ernesto F. Eiroa and Claudio Simeone [arXiv: gr-qc/1008.0382].

[15] Amrita Bhattacharya, Ilnur Nigmatzyanov, Ramil Izmailov, and Kamal K.
Nandi, Class. Quant. Grav. \textbf{26}, 235017 (2009).

[16] G.T. Horowitz and S.F. Ross, Phys. Rev. D \textbf{56}, 2180 (1997).

[17] H. A. Buchdahl, Phys. Rev. \textbf{115}, 1325 (1959). There is a bit of
history here: Bronnikov and Grinyok [Grav. \& Cosmol. \textbf{7}, 297 (2001)]
point out that the solution was first derived by Fisher [Zh. Eksp. Teor. Fiz.
\textbf{18}, 636 (1948)] for a canonical scalar and that it been independently
rediscovered by Bergmann and Leipnik in 1957 (also called anti-Fisher
solution) for a phantom scalar, corresponding to Brans-Dicke $\omega<-3/2$.
Buchdahl mentions that their solution is disfigured by an inconvenient choice
of coordinates.

[18] K.A. Bronnikov, M.V. Skvortsova, and A.A. Starobinsky, Grav.
Cosmol.\textbf{16}, 216 (2010) [arXiv:gr-qc/1005.3262v1].

[19] K.A. Bronnikov and A.A. Starobinsky, JETP Lett. \textbf{85},1 (2007).

[20] Israel Quiros, Rolando Bonal, and Ronaldo Cardenas, Phys. Rev. D
\textbf{62}, 044042 (2000).

[21] I. Quiros, Phys. Rev. D \textbf{61}, 124026 (2000)

[22] H.G. Ellis, J. Math. Phys. \textbf{14}, 104 (1973); \textit{ibid}.
\textbf{15}, 520E (1974). The solution has been independently discovered also
by K.A. Bronnikov, Acta Phys. Polon. B \textbf{4}, 251 (1973). We call it
Ellis-Bronnikov wormhole.

[23] C. H. Brans, Phys. Rev. \textbf{125}, 2194 (1962).

[24] Arunava Bhadra and Kabita Sarkar, Gen. Rel. Grav. \textbf{37}, 2189 (2005).

[25] J.E. Lidsey, D. Wands, and E.J. Copeland, Phys. Rep. \textbf{337}, 343
(2000) [hep-th/9909061v2].

[26] M. Gasperini and G. Veneziano, Phys. Rep. \textbf{373}, 1 (2003) [hep-th/0207130].

[27] C. Armend\'{a}riz-Pic\'{o}n, Phys. Rev. D \textbf{65}, 104010 (2002).

[28] \'{E}.\'{E}. Flanagan, Class. Quant. Grav. \textbf{21}, 3187 (2004).

[29] Y.M. Cho, Phys. Rev. Lett. \textbf{68}, 3133 (1992).

[30] V. Faraoni, Phys. Lett. A \textbf{245}, 26 (1998).

[31] C.H. Brans, Class. Quant. Grav. \textbf{5}, L197 (1988).

[32] M. S. Morris and K. S. Thorne, Am. J. Phys. \textbf{56}, 395 (1988).

[33] S. Weinberg, \textit{Gravitation \& Cosmology}, John Wiley, New York, 1972.

[34] C.M. Will, Living Rev. Relativ. \textbf{4}, 2001-2004 (2001).

[35] D. Hochberg and M. Visser, Phys. Rev. Lett. \textbf{81}, 746 (1998).

[36] T.K. Dey and S. Sen, Mod. Phys. Lett. A \textbf{23}, 953 (2008).

[37] Amrita Bhattacharya and Alexander A. Potapov, Mod. Phys. Lett. A
\textbf{29}, 2399 (2010).

[38] T. Matsuda, Prog. Theor. Phys. \textbf{47}, 738 (1972).

[39] J.A. Gonzalez, F.S. Guzman and O. Sarbach, Class. Quant. Grav.\textbf{
26}, 015010 (2009); \textit{ibid}. \textbf{26}, 015011 (2009).
\end{document}